\documentstyle[12pt,epsf]{article}
\def\bc{\begin{center}}
\def\ec{\end{center}}
\def\be{\begin{equation}}
\def\ee{\end{equation}}
\def\bea{\begin{eqnarray}}
\def\eea{\end{eqnarray}}
\def\ra{\rightarrow}

\def\kmung{$K~\rightarrow~\mu\nu\gamma$}
\def\GF{G_{\mbox{\tiny F}}}
\def\G{G_{\mbox{\tiny F}} e V_{us}^\ast}
\def\eps{\varepsilon^{\mu\nu\alpha\beta}}
\def\xio{\xi_{\mbox{\tiny odd}}}
\def\xiem{\xi_{\mbox{\tiny EM}}}
\def\M{{\cal M}}
\begin{document}
\thispagestyle{empty}
\renewcommand{\thefootnote}{\fnsymbol{footnote}}
\renewcommand{\thesection}{\arabic{section}.}

{\Large \bf A Possibility to Measure $CP$-Violating Effects \\
in the Decay \kmung}\\[2mm]
\bc
R.N. Rogalyov\footnote{E-mail: rogalyov@mx.ihep.su}  \\[2mm]
Institute for High Energy Physics, Protvino, Russia. \\[3mm]
\ec

{\small It is argued that a precise measurement of the transverse component of the muon spin
in the decay \kmung \ makes it possible to obtain more stringent limits
on $CP$-violating parameters of the leptoquark, SUSY and left-right symmetric models.
The results of the calculations of the $CP$-even transverse component of the muon spin 
in the decay \kmung\ due to the final-state electromagnetic interactions are presented.
The weighted average of the transverse component of the muon spin comprises 
$\sim 2.3 \times 10^{-4}$. }

\renewcommand{\thefootnote}{\arabic{footnote}}
\setcounter{footnote}{0}
	
\section{Introduction}

The transverse component of the muon spin in the decay \kmung \ beyond the
Standard Model is due to both the electromagnetic and {\it CP-} and {\it T}-violating
interactions:
\be
\xi = \xiem + \xio,
\ee
where $\xiem$ is the contribution of the electromagnetic Final-State Interactions (FSI)
and $\xio$ is the contribution of the $CP$-odd interactions.

Current limitations on the $CP$-violation parameters in various non-Standard
models allow the transverse component of the muon spin in the decay
\kmung \ to be rather large \cite{Geng}: the left-right symmetric models
based on the symmetry group $SU(2)_L\times SU(2)_R \times U(1)_{B-L}$
with one doublet $\Phi$ and two triplets $\Delta_{L,R}$ of Higgs bosons
can give $\xio \sim 3.5\times 10^{-3}$ \cite{uno}, 
supersymmetric models---$\xio \sim 5\times 10^{-3}$ \cite{Ng}, 
leptoquark models---$\xio \sim 2.5 \times 10^{-3}$ \cite{tre}.
To extract the value of $\xio$ from the experimental data,
one should know the value of $\xiem$ exactly. 

It has long been known that \cite{Khriplovich} the transverse polarization 
of the muon can be accounted for by the imaginary parts of
the form factors parametrizing the expression for the amplitude of the decay.
In this work, we compute the contribution of the electromagnetic FSI to
the transverse component of the muon spin in the decay \kmung \ 
in the one-loop approximation (to be certain, we consider the decay 
$K^+ \ra \mu^+\nu\gamma$). 
Our calculations are performed in the
framework of the Chiral Perturbation Theory (ChPT) \cite{Gasser}.

It should be mentioned that some contributions to $\xiem$ were calculated 
in \cite{Kudenko, Isidori}.
In contrast to the mentioned calculations, we take into account a
complete set of the diagrams contributing to the imaginary part of the decay 
amplitude in the leading order of the ChPT.

For the description of the decay $K^+(p_K) \ra \mu^+(k)\nu(k')\gamma(q)$,
we use the following variables:  
$M_K=494$~MeV and $m_\ell=106$~MeV are the kaon and muon masses;
\bea
&& x=\frac{2p_K\cdot q}{M_K^2};\ \ y =\frac{2p_K\cdot k}{M_K^2};\ \ \ 
\lambda=\frac{1-y+\rho}{x}; \ \
 \rho=\frac{m_\ell^2}{M_K^2}; \ \ \gamma=\frac{F_A}{F_V}; \\ \nonumber
 && \tau\; =\; (1-\lambda)x + \rho; \ \ \ \zeta = 1-\lambda-\tau; \ \ 
 F_V\; =\; {\sqrt{2}\,M_K \over 8\pi^2\,F}; \ \ \ \ 
 F_A\; = \; {4\sqrt{2}\,M_K\, \left(L_9+L_{10}\right) \over F}; \nonumber
\eea
$L_9=6.9\pm 0.7\times 10^{-3}$ and $L_{10}=-5.5\pm 0.7\times 10^{-3}$ are the parameters
of the $O(p^4)$ ChPT Lagrangian; and $F=93$~MeV.
The relevant terms of the ChPT Lagrangian \cite{Gasser, DAFNE} have the form 
\bea \label{LagrCHPTkmung}
L_{\mbox{\tiny CHPT}}^{K\,\rightarrow \mu\nu\gamma}\!\!\!\!\! && \!\!\!\!\! = 
F\GF V_{us} \partial_\mu K^+ l_\mu^- - e\; \bar \mu\hat A \mu 
+ieA_\mu \partial_\mu K^+ K^- 
+i\G F K^- A_\mu l_\mu^+  -\\ \nonumber
&&-\; {\G \over F} \left({1\over 8\pi^2}\; \eps \partial_\mu K^+ \partial_\alpha A_\nu l_\beta^- - 4\sqrt{2}\; iM_\pi(L_9+L_{10})\; \partial_\mu K^+ l_\nu^- (\partial_\mu A_\nu - \partial_\nu A_\mu)\right)- \\ \nonumber
&& - \ {\alpha \over 2\pi F} \eps \partial_\mu A_\nu \partial_\alpha A_\beta \pi^0
+ {i \GF \over 2} V_{us}^\ast l_{\mu}^-
\left( K^+ \partial_\mu \pi^0\; -\; \pi^0 \partial_\mu K^+ \right), \nonumber
\eea
where $\GF = 1.17\times 10^{-5}$~GeV$^{-2}$ is the Fermi constant; 
$\alpha = e^2/(4\pi) = 7.3\times10^{-3}$, $e$ is the electron charge; 
$V_{us} = 0.22$ is the element of the Cabibbo--Kobayashi--Maskawa matrix;
$K^+, \pi^0, A, \nu \ \mbox{and} \ \mu$ are the fields of the $K^+$ meson, 
$\pi^0$ meson, photon, antineutrino, and muon, respectively; 
$l^+_\beta = \bar \mu \gamma_\beta (1-\gamma^5)\nu$; 
$l^-_\beta = \bar \nu \gamma_\beta (1-\gamma^5)\mu$.

\section{ Expression for Polarization of Muon \\
in Terms of Helicity Amplitudes}

Experimentally, the transverse component of the muon spin can be defined as follows:
\be
\xi = {N_+ - N_- \over 2\; (N_+ + N_-)},
\ee
where $N_+$($N_-$) is the number of the produced muons whose spin is directed 
along(against) a beforehand specified direction of polarization. 
We introduce vector $\vec o$ 
specifying such direction in the case under consideration. In the kaon rest frame,
it is orthogonal to the vectors $\vec q, \vec k,$ and $\vec k'$ (in this frame,
these three vectors are linearly dependent):
\be
\vec {\bf o} = {2\over M_K^2\, x\sqrt{\lambda\zeta}} ({\bf \vec q \times \vec k}),
\ee
a positive value of $\xi$ implies that the projection of spin of muon on 
vector $\vec o$ is positive: $\vec s\vec o >0$.

The respective 4-vector is defined as
the unit vector orthogonal to the vectors $q, k,$ and $k'$\footnote{Here and below,
$\epsilon^{0123} = -1, \ \ {\mbox Tr}\;
\gamma^5\gamma^\mu\gamma^\nu\gamma^\alpha\gamma^\beta = 4 i \eps.$}:
\be\label{KmungVectoro}
o^\lambda = {2\over M_K^3\, x\sqrt{\lambda\zeta}}\; \varepsilon^{\mu\nu\rho\lambda}
k'_\mu k_\nu q_\rho,
\ee
or, to put it differently,
\be
o^\mu={\omega_-^\mu(k,k') - \omega_+^\mu(k,k') \over i\sqrt{2}},
\ee
where the vectors $\omega_-^\mu(k,k')$ and $\omega_+^\mu(k,k')$ are defined
by the relations
\bea\label{OmegaKmung}
\hat \omega_+(k,k') &=& - {\sqrt{2} \over 2M_K^3 x \sqrt{\lambda\zeta}}
\left(\hat k \hat q \hat k' (1-\gamma^5) + \hat k' \hat q \hat k (1+\gamma^5) - {2\rho x \lambda M_K^2 \over 1-x-\rho} \hat k' \right), \\ \nonumber
\hat \omega_-(k,k') &=& - {\sqrt{2} \over 2M_K^3 x \sqrt{\lambda\zeta}}
\left(\hat k \hat q \hat k' (1+\gamma^5) + \hat k' \hat q \hat k (1-\gamma^5) - {2\rho x \lambda M_K^2 \over 1-x-\rho} \hat k' \right).  \nonumber
\eea

The helicity amplitudes for the decay $K^+(p) \ra \mu^+(k)\nu(k')\gamma(q)$ 
are defined as follows:
\be
\M_{r\,s} = \langle \mu_s(k) \nu(k')\gamma_r (q) | \M |K(p)\rangle
\ee
where $r=\pm$ is the helicity of the photon; $s=\pm$ is the helicity of the muon in
the reference frame comoving with the center of mass of the muon and neutrino,
and the amplitude $\M$ is defined by
\[
S=1-(2\pi)^4i\delta(k+k'+q-p)\M,
\]
where $S$ is the scattering matrix in the respective channel.

The particles produced in the decay \kmung \ can be described by the
wave function
\bea
\label{WaveFunction}
\hspace*{-4mm}|\Psi\rangle = S|K(p)\rangle & \!\!\!\!=\!\!\!\!& {1 \over \Gamma} \int d\Phi
\left( \M_{-\,-}|\gamma_-(q) \mu_-(k) \nu(k')\rangle \right.
+ \M_{-\,+}|\gamma_-(q) \mu_+(k) \nu(k')\rangle \\ \nonumber
&& + \M_{+\,-}|\gamma_+(q) \mu_-(k) \nu(k')\rangle
\left. + \M_{+\,+}|\gamma_+(q) \mu_+(k) \nu(k')\rangle \right), \nonumber
\eea
where $\Gamma$ is the decay width, and the element of the phase space
has the form
\[
d\Phi = {1\over (2\pi)^5} \delta(k+k'+q-p) {d^3 {\bf k} \over 2 k_0}
{d^3 {\bf k'} \over 2 k'_0}{d^3 {\bf q} \over 2 q_0}.
\]

The operator of spin $s_\mu$ acts on fermion states as follows:
\be
s_\mu={W_\mu \over m} = -\; {\gamma_\mu \gamma^5 \over 2} \hat \varepsilon_0,
\ee
where $W_\mu$ is the Pauli--Lubanski vector and $\hat \varepsilon_0$ is the
operator of the sign of energy. The average value of the transverse component
of spin in the state $|\Psi\rangle$ is equal to $ \langle\Psi|(-\;s_\mu \cdot o_\mu)|\Psi\rangle$,

Since 
\bea
\langle\mu_-(\vec k)|s^\nu |\mu_-(\vec k)\rangle &=&
- {1 \over 4m_\ell} \bar v(k,N) \gamma^\nu \gamma^5 v(k,N) \ \ \ \ \
=  \ \ {N^\nu \over 2},  \\ \nonumber
\langle\mu_-(\vec k)|s^\nu |\mu_+(\vec k)\rangle &=&
- {1 \over 4m_\ell} \bar v(k,-N) \gamma^\nu \gamma^5 v(k,N) \ \ = 
-\; {\omega_-^\nu \over \sqrt{2}}, \\ \nonumber
\langle\mu_+(\vec k)|s^\nu |\mu_-(\vec k)\rangle &=&
- {1 \over 4m_\ell} \bar v(k,N) \gamma^\nu \gamma^5 v(k,-N) \ \  = 
- \; {\omega_+^\nu \over \sqrt{2}}, \\ \nonumber
\langle\mu_+(\vec k)|s^\nu |\mu_+(\vec k)\rangle &=&
- {1 \over 4m_\ell} \bar v(k,- N) \gamma^\nu \gamma^5 v(k,-N) = 
- \; {N^\nu \over 2} ,  \nonumber
\eea
where spinor $v(k,N)$ describes the muon of momentum $k$ and vector of spin $N$,
\be
N_\nu = {(1-x-\rho) k_\nu - 2\rho k'_\nu \over m_\ell (1-x-\rho)},
\ee
the expectation value of the transverse component of the muon spin
is determined by the relation
\bea
\label{TrSpinKmung}
\xi={\Xi \over {\cal N}^2}\equiv{1 \over {\cal N}^2}
\left( \M_{-\; -}' \M_{-\; +}'' - \M_{-\; +}' \M_{-\; -}'' \right.
\left. + \M_{+\; -}' \M_{+\; +}'' - \M_{+\; +}' \M_{+\; -}'' \right),
\eea
where ${\cal N}$ is the normalization factor, ${\cal N}^2=\sum_{i,j=\pm}|\M_{i,j}|^2$;
$\M_{r,s} = \M_{r,s}' + i\M_{r,s}'' \ (r,s=\pm)$
(this formula is readily obtained by isolating an infinitesimal volume
of the phase space of the particles produced in the decay and employing
formula (\ref{WaveFunction})).

In the calculations of the helicity amplitudes we use the so called
diagonal spin basis \cite{Galynskii, Sikach, CORE} formed by the vectors $\omega^\mu_\pm$
and light-like linear combinations of the vectors $k$ and $k'$.

With the use of this basis, the helicity amplitude  $\M_{r,s}$ 
can be represented in a manifestly covariant form
\be
\label{HelAmplKmung}
\M_{r,s}=\bar u(k')\M_\alpha (k,k',q) \epsilon_\alpha(r) v(k,sN) 
= Tr \M_\alpha (k,k',q) \epsilon_\alpha(r) v(k,sN) \bar u(k'),
\ee
where the expression for $\M_\alpha (k,k',q)$ is given by the Feynman diagrams,
the polarization vectors of the photon are equal to
\be\label{EpsilonKmung}
\epsilon_\mu(\pm) = {\sqrt{2} \over 2M_K x \sqrt{\lambda\zeta}}
\left(-x\lambda k_\mu+x(1-\lambda)k'_\mu -(1-\rho-x)q_\mu \mp i\varepsilon_{kk'q\mu}\right),
\ee
and the quantities $v(k,sN) \bar u(k')$ can be brought in the form
\bea \label{BilinCombKmung}
v_\mu(k,-N) \bar u_\nu(k') &=& {(\hat k - m_\ell) \hat k' \over	2 M_K \sqrt{1-x-\rho}} 
(1+\gamma^5), \\ \nonumber
v_\mu(k,N) \bar u_\nu(k') \ &=& {M_K^2 (1-x-\rho) - m_\ell \hat k'  \over
2 M_K \sqrt{1-x-\rho}} \; \hat \omega_- (1+\gamma^5). \nonumber
\eea

The leading contribution to the real part of the decay amplitude is given by 
the tree diagrams corresponding to the Lagrangian (\ref{LagrCHPTkmung}) \cite{DAFNE} (see Fig.~1).
 The helicity amplitudes for the decay $K^+ \ra \mu^+ \nu \gamma$
in the tree approximation have the form 
\bea \label{HelAmpliKmungYaf}
\M_{-\;-} &=& 2\,i\G m_\ell x 
\sqrt{\lambda \zeta \over 1-x-\rho}
\left({\sqrt{2} F (1-\rho)\over x^2 (1-\lambda)} - \right.
\left. M_K {F_V-F_A \over 2} \right), \\ \nonumber
\M_{-\;+} &=& -2\,i\G {x\lambda \over \sqrt{1-x-\rho}}
\left(m_\ell F {\sqrt{2\rho} \over x(1-\lambda)} - {F_V-F_A \over 2} M_K^2 (1-x) \right),\\ \nonumber
\M_{+\;-} &=& 2\,i\G m_\ell x 
\sqrt{\lambda \zeta \over 1-x-\rho}
\left(F {\sqrt{2} (1-x-\rho) \over x^2 (1-\lambda)} \right.
\left. + {F_V+F_A \over 2} M_K \right), \\ \nonumber
\M_{+\;+} &=& i\G { (F_V+F_A )x\over \sqrt{1-x-\rho}}
 M^2_K \zeta,  \nonumber
\eea
where the first index in the left-hand side denotes the polarization of the photon
and the second---the polarization of the muon in the reference frame comoving with the
center of mass of the lepton pair.
The calculation of the imaginary parts of the helicity amplitudes is considered in the 
following Section.

The differential probability for the decay is determined by the matrix element squared
\bea\label{KmunfMesqYaf}
\sum_{polariz.}|{\cal M}|^2 = |\G|^2\;\left(m_\ell^2F^2\,IB \,+\,
{(F_V+F_A)^2 \over 2M^2_K} SD_+ + {(F_V-F_A)^2 \over 2M^2_K} SD_- \right. + \\ \nonumber
+ \left. m_\ell F\,{F_V+F_A \over \sqrt{2}M_K} INT_+ + m_\ell F\,{F_V-F_A \over \sqrt{2}M_K} INT_- \right),
\eea
where
\bea
IB &=& {8\lambda \over x^2 (1-\lambda)} \ 
\left( x^2+2(1-x)(1-\rho)-{2\rho(1-\rho) \over 1-\lambda } \right), \\ \nonumber
SD_+ &=& 2 M^6_K x^2 (1-\lambda) \zeta, \\ \nonumber
SD_- &=& 2 M^6_K x^2 \lambda \left( (1-x)\lambda + \rho \right), \\ \nonumber
INT_+ &=& {8 M^2_K m_\ell \lambda \over 1-\lambda} \zeta, \\ \nonumber
INT_- &=& -\;{8 M_K^2 m_\ell \lambda \over 1-\lambda} \left( 1-\lambda + \lambda x - \rho \right). \nonumber
\eea

\section{ Contribution of FSI to Imaginary Part \\  of Decay Amplitude}
The imaginary part of the amplitude for the decay \kmung \ in the leading
order of the perturbation theory is described by the diagrams in Fig.~2.
We take into account the diagrams in Figs.~2$g,h$ omitted by the authors of
\cite{Kudenko} in spite of the fact that they are of the same order of magnitude.

We employ the Cutkosky rules \cite{Zuber} to replace the propagators 
with the $\delta$ functions. Thus we obtain the expression for the
imaginary part of the amplitude in terms of the integrals:
\be
\label{InteKmung}
\M''_i=-\, {\alpha F \over 2\pi}\G \int dr {\Delta \over N_i(r\cdot q, r\cdot k)} 
\bar u(k')(1+\gamma^5)T_i(r,k,k',q,\epsilon)v(k),
\ee
where $N_i$ is the product of the remaining propagators in the respective diagram and
$T_i$ are the respective tensor structures (label $i$ specifies the diagram
in Fig.~2, $i=$a$\div$i); in the case of the diagrams in Figs.~2$a$--$h$, 
$\Delta = \delta(r^2-m_\ell^2) \delta \left((k+q-r)^2\right)$
whereas, for the diagram in Fig.~2$i$ $\Delta = \delta((r+q)^2-M_\pi^2)\delta((k-r)^2-m_\ell^2)$ 
($M_\pi = 135$~MeV --- is the mass of the $\pi^0$ meson). 

The computations of the diagrams in Fig.~2 are made with the {\it REDUCE} package.
These diagrams are calculated exactly, no approximation is used.

The calculated imaginary part of the amplitude of the decay \kmung\ takes the form
\be\label{TheKmungResult}
{\cal M''} =\;-\;{\G \over 4\pi} \bar u_\nu(k') (1+\gamma^5) 
\left( {\cal M}^{IB} + {\cal M}^{SD} + {\cal M}^{(\pi)} \right) v_\mu (k),
\ee
where
\be
{\cal M}^{IB} = {2 \pi \alpha F\over M_K^2} \sum_{n=1}^4 c_n^{IB} {\cal E}_n 	 \nonumber
\ee
--- is the contribution of the diagrams in Figs. 2$a$, 2$b$, 2$c$, 2$d$, 2$g$, 2$h$;
\be
{\cal M}^{SD} = {\pi \sqrt{2} \alpha \over M_K}
\sum_{n=1}^4 (-F_A c^A_n+F_V c^V_n) {\cal E}_n 	 \nonumber
\ee
--- is the contribution of the diagrams in Figs. 2$e$ and 2$f$ and
\be
{\cal M}^{(\pi)} =  {\alpha \over 4 \pi  F} 
\left(c^{(\pi)}_2 {\cal E}_2 +c^{(\pi)}_4 {\cal E}_4\right) \nonumber
\ee
--- is the contribution of the diagram in Fig. 2$i$. 
Tensor structures ${\cal E}_i$ have the form
\bea\label{BaseStructuresKmung}
{\cal E}_1 &=& M_K^2 m_\ell x \left[(1-\lambda) k'\cdot \epsilon - \lambda k\cdot \epsilon \right], \\ \nonumber
{\cal E}_2 &=& M_K^2 \left[ k\cdot \epsilon \hat q - \frac{M_K^2}{2} x(1-\lambda) \hat \epsilon \right], \\ \nonumber
{\cal E}_3 &=& M_K^2 \left[ k'\cdot \epsilon \hat q - \frac{M_K^2}{2} x\lambda \hat \epsilon \right], \\ \nonumber
{\cal E}_4 &=& M_K^2 m_\ell \hat q \hat \alpha,  \nonumber
\eea
and the coefficients in the above expressions are given by 
\bea \label{CoeffCIB}
c_1^{IB} &=& -\; {4 \over (1-\lambda) x} (G_3-(1+\tau) G_2+\rho (F_1-F_2)),\\ \nonumber
c_2^{IB} &=& {4 \rho \over (1-\lambda) x} (2 G_1+(1+\tau) G_2-(1-\tau) G_3-(\tau+\rho) F_1)+2 F_5 \rho, \\ \nonumber
c_3^{IB} &=& 4 \rho (-F_2-G_4),			\\ \nonumber
c_4^{IB} &=& {2 \lambda \over (1-\lambda)} (G_3-G_2-\rho F_2)-2 (G_2+2 G_1-F_1) \\ \nonumber
& & +{4-x (1-\lambda) \over (1-\lambda) x} 
(2 G_1+G_2-(1-\tau) G_3-\rho F_3)  + {2\over (1-\lambda)} (-\tau F_1+\rho F_3)-F_4+F_5 \rho, \nonumber
\eea
\bea \label{CoeffCSD}
c^V_1&=&({1 \over 3} x (1-\lambda)-2 \tau) F_5+(\tau+\rho) F_6, 		\\ \nonumber
c^V_2&=&{1\over 3} (\tau (1+5\tau-14\rho)-\rho (1-3\rho +x\lambda )) F_5
	+\rho (\lambda x+2 \rho) F_6+(1-\tau) F_7- {(1+\lambda) \over (1-\lambda)} F_8,	\\ \nonumber
c^V_3&=&-x (1-\lambda) (\tau+{\rho \over 3}) F_5-\tau F_7+F_8,	\\ \nonumber
c^V_4&=&{1 \over 2} (x (x (1-\lambda)^2+\tau (3-2 \lambda)) F_5
	+x (1-x-\lambda+\lambda x+\rho (3 \lambda-4)) F_6+(1-\tau) F_7),\\ \nonumber
c^A_1&=&c^V_1,		\\ \nonumber
c^A_2&=&c^V_2+2 (-({5 x^2 (1-\lambda)^2 \over 3}-\rho^2) F_5
	+\rho (x-x \lambda-\rho) F_6+\tau F_7),\\ \nonumber
c^A_3&=&c^V_3,		\\ \nonumber
c^A_4&=&c^V_4+{1 \over 2} (-x (1-\lambda) (\tau+2 x (1-\lambda)) F_5
	+4 \rho x (1-\lambda) F_6+3 \tau F_7),	 \nonumber
\eea
\bea\label{CoeffCpion}
c^{(\pi)}_2&=&{1 \over 4 M_K^2\, x^2 (1-\lambda)^2} \theta\left(x-{\kappa+\sqrt{2\kappa\rho} \over 1-\lambda }\right)\;
\left({2 \kappa^2 \rho \over x (1-\lambda)} S_4 + \right. \\ \nonumber
&&	+ ((x^2 (1-\lambda)^2-\rho \kappa) ({x (1-\lambda) \over \tau}+2) 
\left. 	+ x^2 (1-\lambda)^2) {S \over \tau}\right),\\ \nonumber
c^{(\pi)}_4&=&{1 \over 4 M_K^2\, x^2 (1-\lambda)^2} \theta\left(x-{\kappa+\sqrt{2\kappa\rho} \over 1-\lambda }\right)\;
\left({\kappa^2 (2 \tau+\rho) \over x (1-\lambda)} S_4 + \right. 		\\ \nonumber
&&  	+((x^2 (1-\lambda)^2-\rho \kappa) ({x (1-\lambda) \over \tau}+3)
\left. 	-3 \kappa (x (1-\lambda)+\tau)) {S \over 2  \tau} \right), \nonumber
\eea
where 
\be
S =\sqrt{((1-\lambda) x-\kappa)^2-4 \kappa \rho} \, \ \ \ \ \ \ 
\kappa={M_\pi^2 \over M_K^2};
\ee
 $\theta$ function in formula (\ref{CoeffCpion}) isolates the kinematic domain in which
the imaginary part of the diagram in Fig.~2$i$ does not vanish;
\bea\label{KmungFunctionsS}
S_1&=&\ln\left[1+{(1-\lambda) x \over \rho}\right],		\\ \nonumber
S_2&=&\ln[\rho],				\\ \nonumber
S_3&=&\ln\left[{1 - \lambda x + \rho + \sqrt{R} \over 1 - \lambda x + \rho - \sqrt{R}}\right],\\ \nonumber 
S_4&=&\ln\left[{(1-\lambda) x (\kappa-(1-\lambda) x+S)+2 \kappa \rho \over (1-\lambda) x (\kappa-(1-\lambda) x-S)+2 \kappa \rho}\right],\\	\nonumber
R&=&(1-\lambda x+\rho)^2-4 \rho,	 \nonumber
\eea
\bea\label{KmungFunctionsF}
F_0&=&{1 \over 2 M_K^2 (1-\tau) x} 
\left(1-{\rho \over \tau}+{1-\rho \over 1-\tau} (S_1+S_2)\right),	\\ \nonumber
F_1&=&{1 \over 4 M_K^2 x \zeta} 
\left(-2 S_1-S_2+ {1-\lambda x+\rho \over \sqrt{R}} S_3-{2 \rho \over (1-\lambda) \sqrt{R}} S_3\right),	\\ \nonumber
F_2&=&{1 \over 4 M_K^2 \lambda x \zeta} 
\left({2 \lambda \over 1-\tau} S_1-{\zeta-\lambda \over 1-\tau} S_2-{\zeta-\lambda+x \over \sqrt{R}} S_3\right),	\\ \nonumber
F_3&=& {1\over 2 M_K^2 (1-\lambda) x \sqrt{R}} S_3,			\\ \nonumber
F_4&=& {1\over M_K^2 (1-\lambda)^2 x^2} ((1-\lambda) x-\rho S_1),		\\ \nonumber
F_5&=& {1\over 2 M_K^2 \tau^2},							\\ \nonumber
F_6&=& {1\over M_K^2 x (1-\lambda)} 
\left({S_1 \over x (1-\lambda)} - {1 \over \tau}\right),		\\ \nonumber
F_7&=& {-x (1-\lambda) \over 6 M_K^2 \tau^3}  (x-x \lambda+3 \rho),		\\ \nonumber
F_8&=& {\rho \over M_K^2 x (1-\lambda)} 
\left({x-x \lambda+2 \rho \over x (1-\lambda)} S_1-2\right),	\nonumber
\eea
\bea\label{KmungFunctionsG}
	G_1&=&{1 \over 8 M_K^2 \lambda x \zeta} 
\left({2 \lambda \over (1-\tau)} (\rho-\tau^2) S_1 + \right. \\ \nonumber
&& \left. +{1-\lambda \over 1-\tau } (1-2 \rho-x-\tau x+\tau^2) S_2+(1-\lambda) \sqrt{R} S_3\right), \\ \nonumber
	G_2&=&{1\over\zeta} (-\lambda \rho F_3+2 \lambda G_1-(1-\tau) F_0),\\ \nonumber
	G_3&=&{1 \over \zeta} (-\rho F_3+2 G_1-F_0),	\\ \nonumber
	G_4&=&{1 \over \lambda x} (-2 G_1+\tau (G_2-F_1)+\rho (F_3-F_1)). \nonumber
\eea

Substituting the expressions (\ref{TheKmungResult})--(\ref{CoeffCpion})
in formula  (\ref{TrSpinKmung}), we represent the transverse muon polarization
in the form
\be \label{TrSpinKmung1}
\xiem = -\; \frac{\sum_{n=1}^4 c_n Y_n}{\sum_{r,s=\pm}|\M_{r,s}|^2},
\ee
where
\be
c_n={\alpha\over 4}\; {\G \over M_K^2}
\left(2F c_n^{IB}+\sqrt{2}\, M_K (c_n^V F_V-c_n^A F_A)+{M_K^2\over 4\pi^2 F} c_n^{(\pi)}\right),
\ee
\bea
Y_n &=&\bar u(k')(1+\gamma^5) {\cal E}_n^\alpha
\left(\epsilon_\alpha^-(q)(\M'_{-,-}v_+(k)-\M'_{-,+}v_-(k)) +\right. \\ \nonumber
&& \hspace*{20mm}\left.+\epsilon_\alpha^+(q)(\M'_{+,-}v_+(k)-\M'_{+,+}v_-(k)) \right),
\eea
$v_\pm(k)=v(k,\pm N)$. Since the imaginary parts of the amplitudes 
under consideration are much less than
the respective real parts ($\M''_{r,s} <\!\!\!< \M'_{r,s}$), the denominator of the expression
(\ref{TrSpinKmung1}) is determined by the equation (\ref{KmunfMesqYaf}).
The coefficients $c_n^{IB}, c_n^V, c_n^A, c_2^{(\pi)},$ and $c_4^{(\pi)}$ 
are given in formulas (\ref{CoeffCIB})--(\ref{CoeffCpion}); $c_1^{(\pi)}=c_1^{(\pi)}=0$; 
and
\bea
Y_1&=& {\G m_\ell M_K^3  \sqrt{2\lambda\zeta} \over 1-\lambda} \left( M_K x^2(1-\lambda)\left((F_V-F_A)(1-x-\rho) + \right.\right. \\ \nonumber
&& + \left.\left. 2\,F_A\zeta\right)-2\sqrt{2}F\rho\lambda \right), \\ \nonumber
Y_2&=&{\G m_\ell M_K^3  \sqrt{2\lambda\zeta} \over 1-\lambda} \; \left( M_K x^2\lambda (1-\lambda)(F_V-F_A) +2\sqrt{2}F (-\zeta+\lambda\rho ) \right), \\ \nonumber
Y_3&=& {\G m_\ell M_K^3  \sqrt{2\lambda\zeta} \over 1-\lambda} \left( M_K x^2\lambda (1-\lambda)(F_A-F_V)+2\sqrt{2}F \lambda(1-\rho)  \right), \\ \nonumber
Y_4&=& {\G m_\ell M_K^3  \sqrt{2\lambda\zeta} \over 1-\lambda} 
\left( 2M_K x^2\lambda (1-\lambda)(F_A-F_V) -4\sqrt{2}F \lambda\rho \right). \nonumber
\eea

\section{Discussion of Results and Conclusion}
The transverse component of the muon spin in the decay \kmung \ is plotted in Figs.~3 and~4
as a function of the kinematic variables $x$ and $y$. As is seen, it varies through the range 
$(0\div 7)\times 10^{-4}$ and the the weighted average is equal to 
(the notation see in formula (\ref{TrSpinKmung})
\bea\label{eq:WeghtedAverage}
\langle\xiem\rangle = {\int_{x_{min}} dx \int dy \Xi \over \int_{x_{min}} dx \int dy {\cal N}^2}
\simeq 2.3\times 10^{-4},
\eea
where the lower limit of the integration with respect to $x$, $x_{min}=0.1$,
corresponds to the cutoff energy of the photon $\simeq 25 MeV$.
The accuracy of the result $\simeq 20\%$ is determined by the accuracy of 
the ChPT in order $O(p^4)$ at these energies.
Note that $\xiem$ is negative in sign over all Dalitz plot (positive 
direction is given by the vector $\vec o$ introduced in Section 2.)

The values of the parameters $F_V$ and $F_A$ used in our plots are:
$F_A=0.042$ and $F_V=0.095$; these values predicted by CHPT
coincide with those used in \cite{Kudenko, Likhoded}.

The transverse polarization (which is twice the muon spin)
agrees well with the results presented recently \cite{Likhoded} and disagree
with \cite{Kudenko} and \cite{Isidori}. The point is that the authors
of \cite{Kudenko, Isidori} took into account only a part of the diagrams contributing
to the transverse polarization of the muon. Our results show that the
diagram estimated in \cite{Isidori} does not give a leading contribution
to the imaginary part of the amplitude and the maximum value of the transverse 
polarization of the muon is overestimated in \cite{Kudenko} by an order of magnitude.
However, it should be emphasized that our results sustantiate
the conclusion made in \cite{Isidori}: "An experimental evidence
of $P_T=2\xi$ at the level of $10^{-3}$ would be a clear signal of physics beyond the SM,"
--- in spite of the fact that the analysis performed in \cite{Isidori} is incomplete.
Our results contradict to the conclusion of \cite{Kudenko}.

Thus an observation of the transverse spin of the muon of the order
$10^{-3}$ in the experiments \cite{Kudenko1, Obraztsov, Abe} would signal 
$CP$ and $T$ violation because the background $CP$-even effect 
does not exceed $7\times 10^{-4}$ and its average value is not over
$3\times 10^{-4}$.
Experiments of this sort can be a good tool for testing the above-mentioned 
non-Standard models.\\

{\it Acknowledgment:} \ I am grateful to A.E.~Chalov, V.V.~Braguta, 
and A.A.~Likhoded for the interest in the study.

\newpage
\baselineskip 17 pt
\begin{figure*} \hbox{
\hspace*{-5pt}
       \epsfxsize=350pt \epsfbox{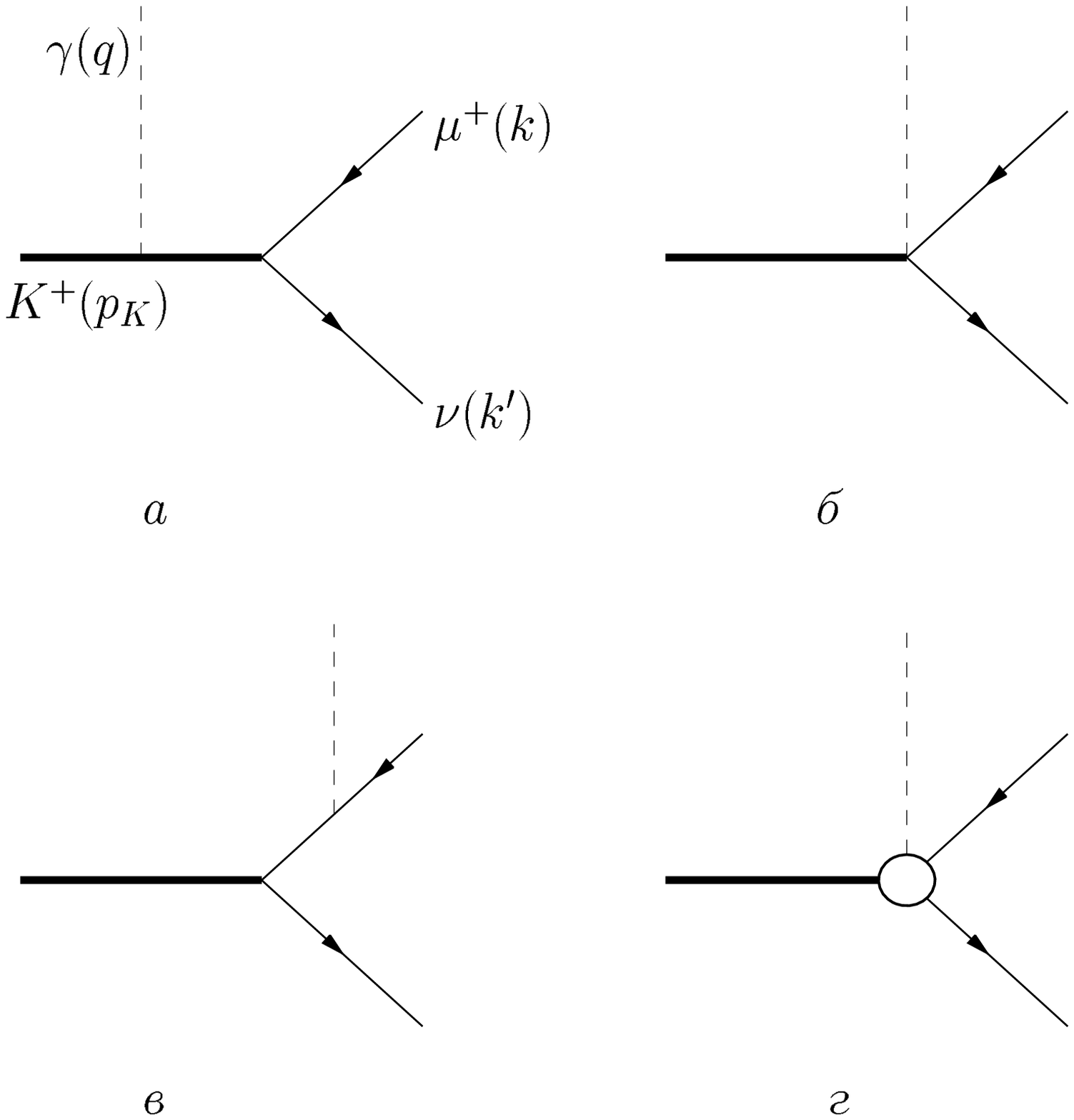} \hspace*{10pt}
       }
\caption{Diagrams describing the decay \kmung \ in the tree approximation.} \label{fig:1}
\end{figure*}

\newpage
\begin{figure*} \hbox{
\hspace*{-5pt}
       \epsfxsize=450pt \epsfbox{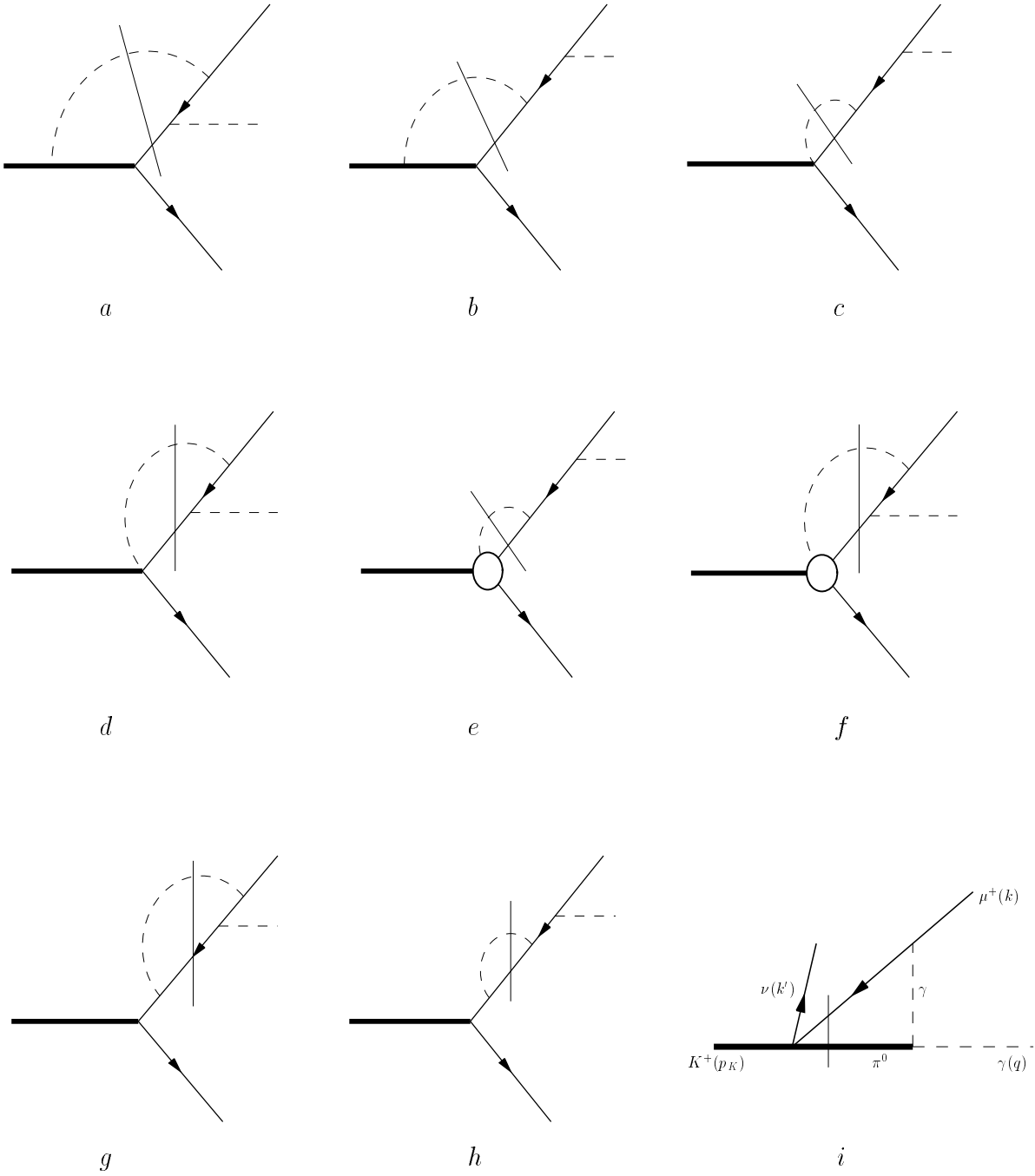} \hspace*{10pt}
       }
\caption{Diagrams giving a contribution to the imaginary part of the 
amplitude of the decay \kmung.} \label{fig:2}
\end{figure*}

\newpage
\begin{figure*} \hbox{
\hspace*{-5pt}
       \epsfxsize=310pt \epsfbox{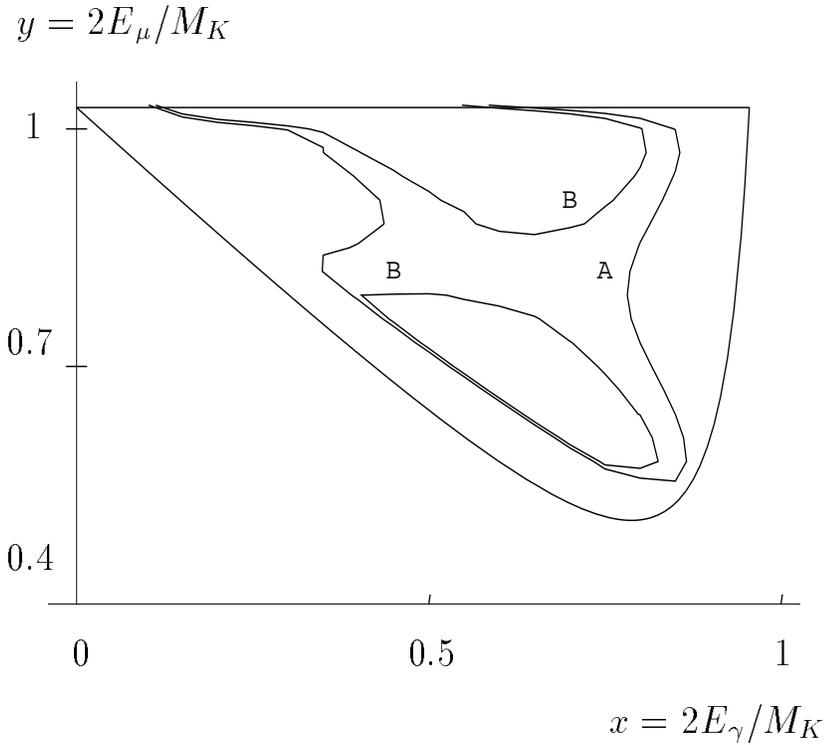} \hspace*{10pt}
       }
\caption{Contour plot for the transverse spin $\xiem$.
Curve {\tt A}: $\xiem =\, 2.5\times 10^{-4}$,
curve {\tt B}: $\xiem =\, 5\times 10^{-4}$.} \label{fig:3}
\end{figure*}

\begin{figure*} \hbox{
\hspace*{-5pt}
       \epsfxsize=310pt \epsfbox{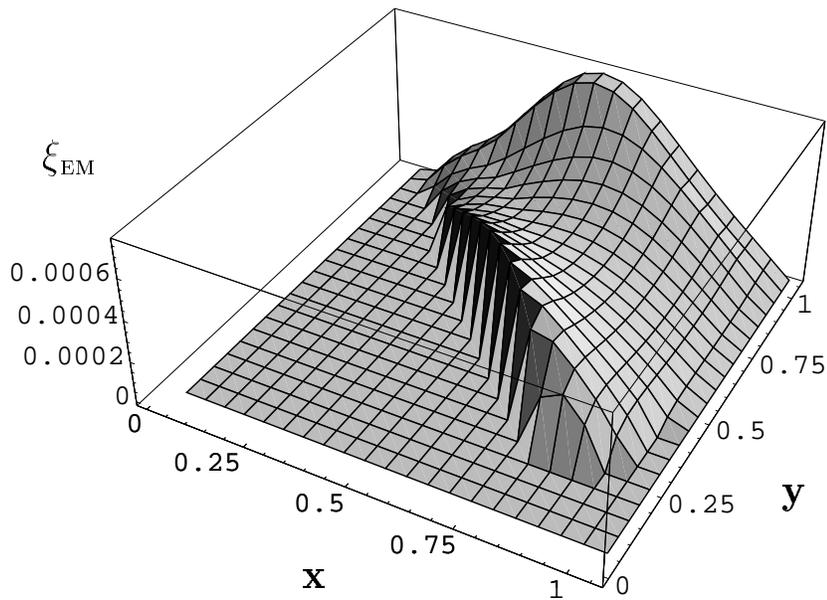} \hspace*{10pt}
       }
\caption{The electromagnetic contribution to the transverse component of the muon spin
over the Dalitz plot for the decay \kmung .} \label{fig:4}
\end{figure*}

\end{document}